\documentclass[12pt]{article}
\let\section=\subsection     \let\subsection=\subsubsection                
\usepackage{graphicx}

\begin{document}

\begin{center}
  {\large \bf CENTRALITY DEPENDENCE OF THERMAL}\\[2mm]
  {\large \bf PARAMETERS IN HEAVY ION COLLISIONS}\\[2mm]
  {\large \bf AT SPS AND RHIC}\footnote{Supported by BMBF 06DR921.}\\[5mm]
  B.~K\"AMPFER$^a$, J.~CLEYMANS$^b$, K.~GALLMEISTER$^c$, S.~WHEATON$^b$\\[5mm]
  {\small \it  
  $^a$ Forschungszentrum Rossendorf, PF 510119, 01314 Dresden, Germany \\
  $^b$ University of Cape Town, Rondebosch 7701, Cape Town, South Africa\\
  $^c$ Institut f\"ur Theoretische Physik, Universit\"at Giessen, 
       Giessen, Germany\\[8mm] }

\end{center}

\begin{abstract}\noindent 
We analyze the centrality dependence of thermal parameters
describing hadron multiplicities, hadron spectra and dilepton
spectra in heavy-ion collisions at SPS and RHIC energies.
\end{abstract}

\section{Introduction} 

It has been shown that various observables 
of relativistic heavy-ion collisions
can be well described by statistical-thermal or hydrodynamical models.
In such a way, a selected subset of a large number of observables
can be reproduced by a small number of characteristic parameters,
such as temperature, density or flow velocity.
It is the subject of the present note to pursue this idea
and to analyze the centrality dependence of the thermal
parameters describing hadron multiplicities, hadron spectra and dilepton spectra.
This will provide further information 
about the effects of the size of the excited strongly interacting
system and help in the systematic understanding of the experimental data. 

\section{Hadron Multiplicities} 

Hadron multiplicities can be reproduced 
\cite{abundance,Becattini} by
the grand-canonical partition function
${\cal Z}_i (V, T, \mu^\alpha) = \mbox{Tr} \left[
\exp\{- (\hat H - \mu^\alpha Q_i^\alpha) / T \} \right]$,
where $\hat H$ is the statistical operator of the system,
$T$ denotes the temperature, and $\mu^\alpha$ and 
$Q_i^\alpha$ represent the 
chemical potentials and corresponding conserved charges
respectively.
The net-zero strangeness and total electric charge constrain
the components $\mu^\alpha$. 
The particle numbers are accordingly 
\begin{equation}
N_i^{\rm prim} = V (2J_i + 1) \int 
\frac{d^3 p}{(2\pi)^3} \, dm_i \,
\left[ \gamma_s^{-S_i}
\mbox{e}^{\frac{E_i - \mu^\alpha Q_i^\alpha}{T}} \pm 1 \right]^{-1}
\mbox{BW} (m_i),
\end{equation}
where we include phenomenologically a strangeness 
saturation factor $\gamma_s$ (with $S_i$ the total number of strange
quarks in hadron species $i$) to account for 
incomplete equilibration in this sector, 
$E_i = \sqrt{\vec p^{\, 2} + m_i^2}$;
$\mbox{BW}$ is the Breit-Wigner distribution
(to be replaced by a $\delta$ function for stable hadrons). 
The final particle numbers are
$N_i = N_i^{\rm prim} + \sum_j \mbox{Br}^{j \to i} N_j^{\rm prim}$
due to decays of unstable particles with branching ratios 
$\mbox{Br}^{j \to i}$.  
Such a description can be justified for multiplicities
measured over the whole phase-space, since many dynamical effects 
cancel out in ratios of hadron yields.

We have analyzed two data sets:
(i) NA49 $4\pi$ multiplicities of
$\langle \pi \rangle = \frac12 (\pi^+ + \pi^-)$, 
$K^\pm$, $\bar p$, $\phi$, and $N_{\rm part}$ 
(taken as the sum over all baryons)
in 6 centrality bins in the reaction Pb(158 AGeV) + Pb 
\cite{Sikler,Blume}
(it should be emphasized that protons are not included
in our analysis \cite{Spencer} as they are not 
participants in non-central collisions), and
(ii) PHENIX mid-rapidity densities of $\pi^\pm$, $K^\pm$, and $p^\pm$
in the reaction Au + Au at $\sqrt{s} = 130$ AGeV in 5 centrality bins
\cite{PHENIX}.
Results of our fits are displayed in fig.~1.
Note that our model tends to underestimate the NA49 $\phi$ yields
for peripheral collisions.
\begin{figure}[th]
\centering
~\\[.1cm]
\includegraphics[width=5cm]{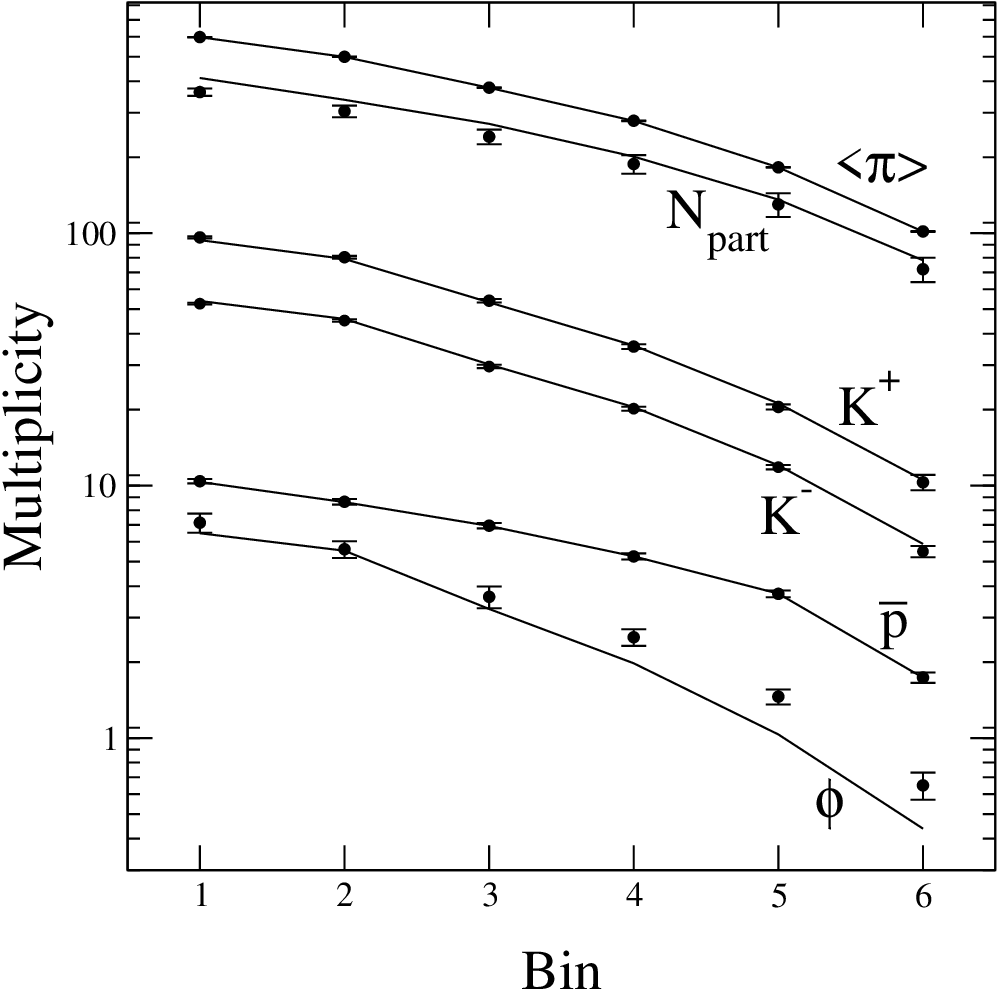}
\hspace*{6mm}
\includegraphics[width=5cm]{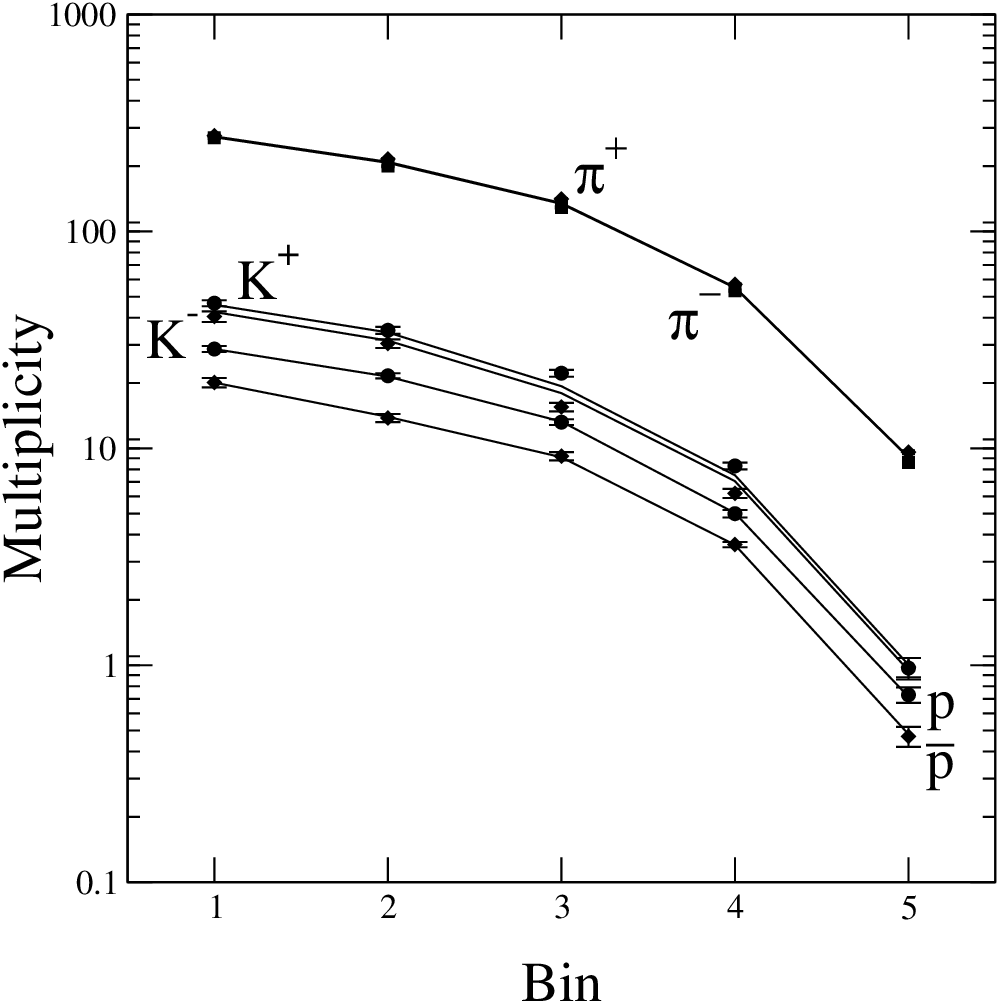}
~\\[.1cm]
\begin{minipage}[t]{14cm}
{\footnotesize Fig.~1:
Comparison of NA49 data (left panel, symbols, \cite{Sikler,Blume})
and PHENIX data (right panel, symbols, \cite{PHENIX})
with our model (lines).
}\end{minipage}
\label{f_1}
\end{figure}

A comparison of the individual thermal parameters of both data sets
is displayed in fig.~2. Most remarkable is the drop
of the baryo-chemical potential $\mu_B$ and the rise
of the strangeness saturation factor $\gamma_s$
when going from $\sqrt{s} = 17$ AGeV to 130 AGeV.
The parameter $\mu_B$ is fairly independent of the centrality.
$T$ seems to stay constant for  $\sqrt{s} = 17$ AGeV, while
at 130 AGeV it rises with $N_{\rm part}$.
The strangeness saturation factor and, of course, 
the volume-equivalent system size
increase significantly with centrality. 
(Note the fiducial meaning
of the volume for the mid-rapidity data of PHENIX.)
Despite the rather limited set of analyzed hadron species,
the extracted thermal parameters describe other hadron yields,
which are at our disposal in central collisions, fairly well. 
\begin{figure}
\centering
~\\[.01cm]
\includegraphics[width=5cm]{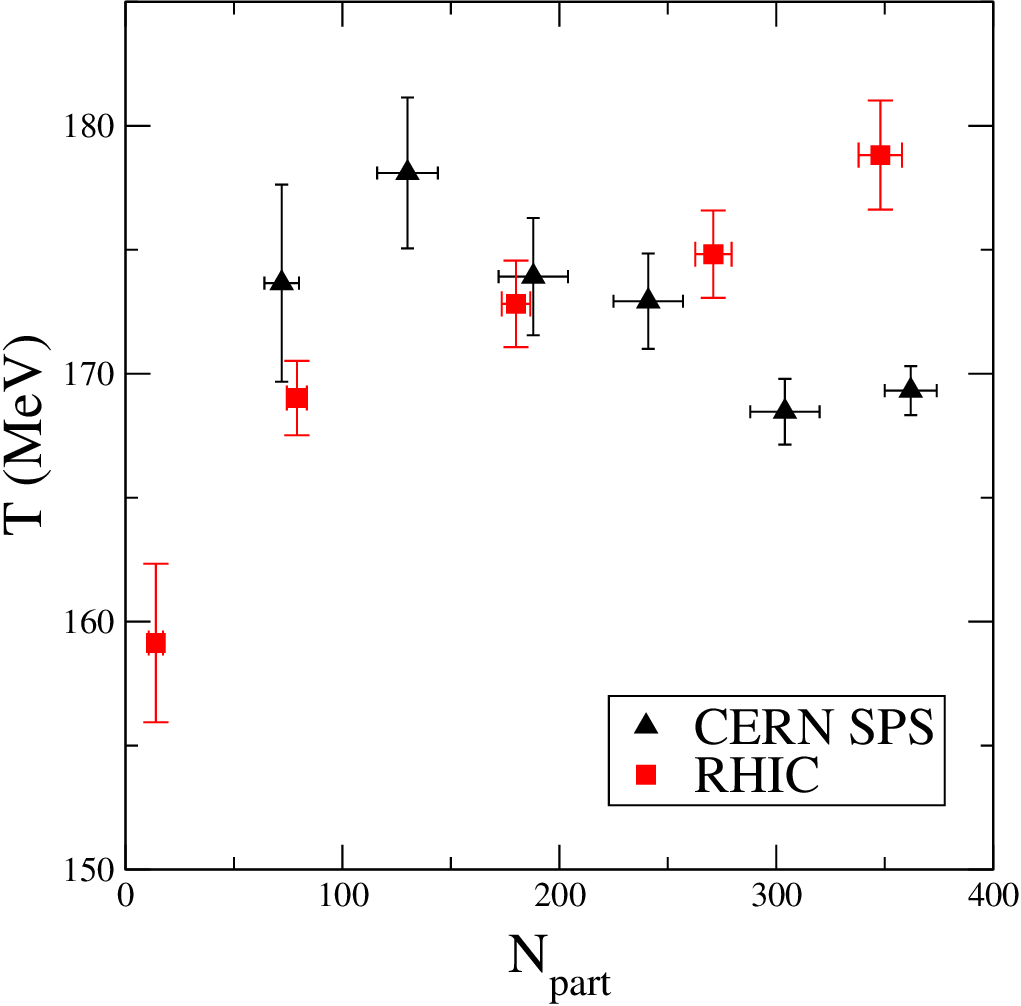}
\hspace*{6mm}
\includegraphics[width=5cm]{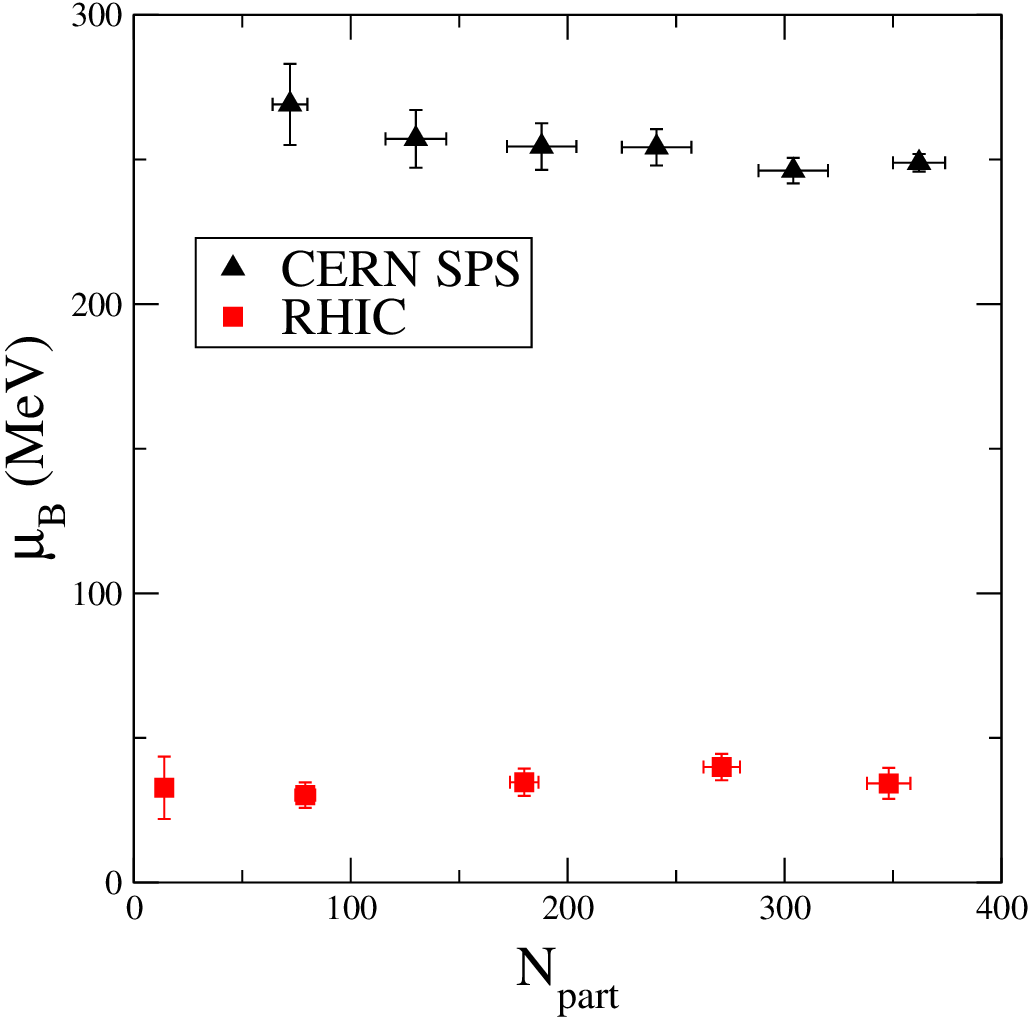}\\[3mm]
\includegraphics[width=5cm]{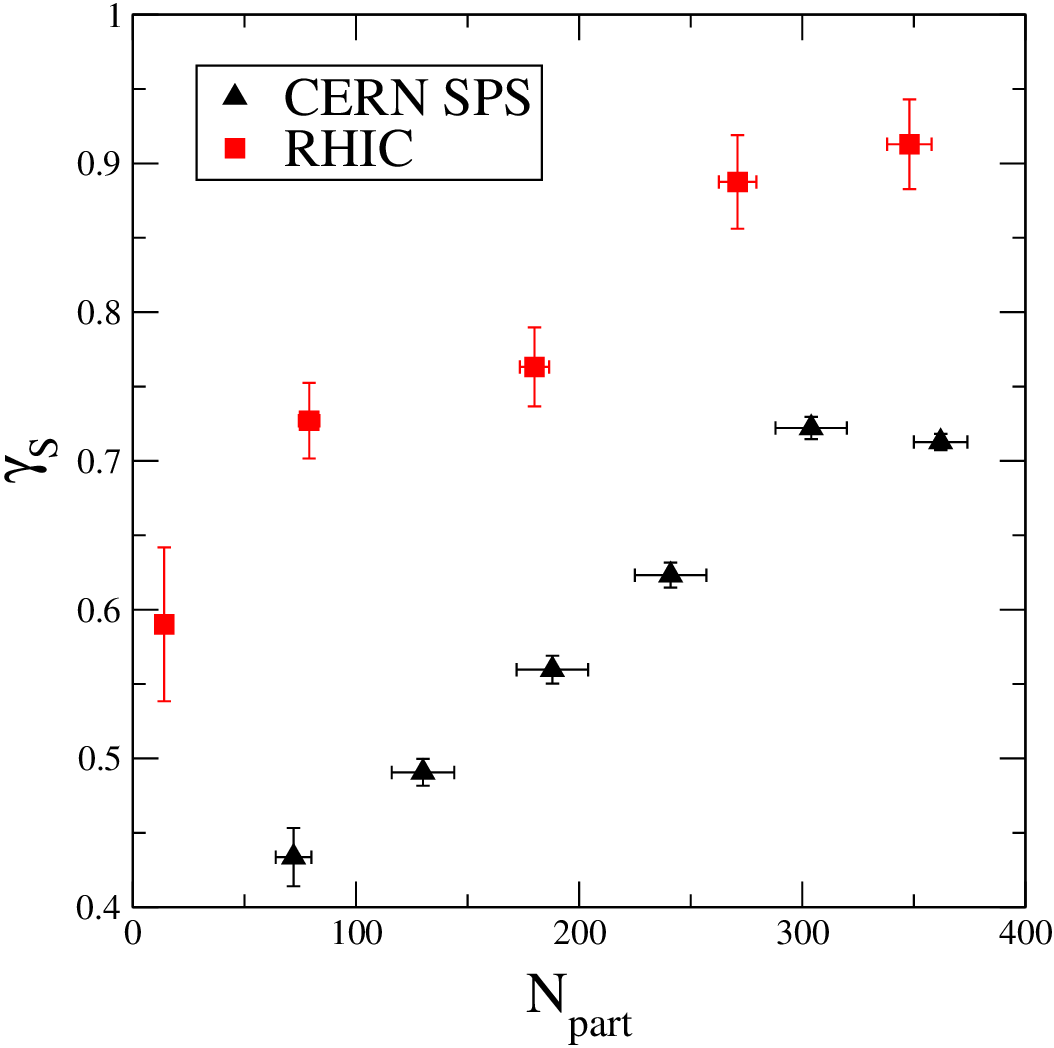}
\hspace*{6mm}
\includegraphics[width=5cm]{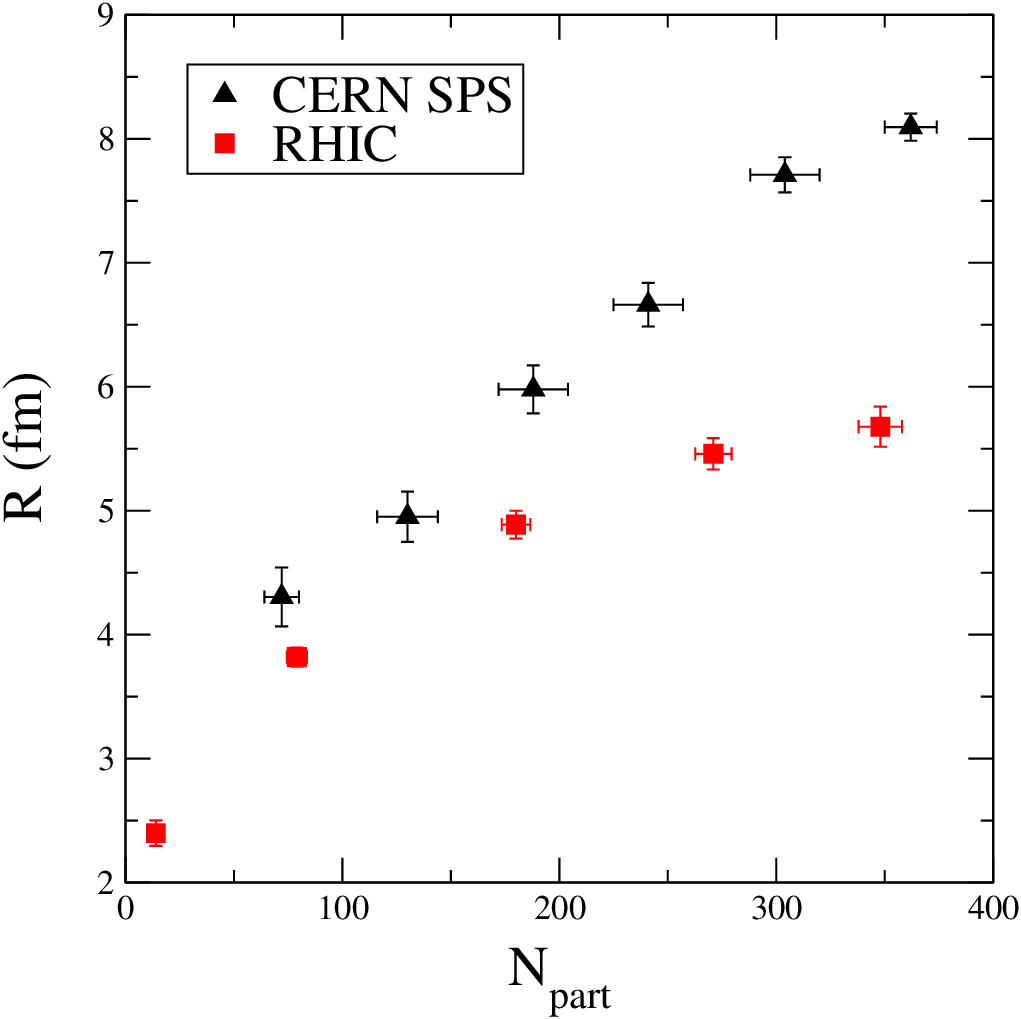}
~\\[.1cm]
\begin{minipage}[t]{14cm}
{\footnotesize Fig.~2:
Temperature, baryo-chemical potential, strangeness saturation factor
and volume-equivalent radius as a function of $N_{\rm part}$.
Triangles (squares) are for the NA49 (PHENIX) 
data \cite{Sikler,Blume} (\cite{PHENIX}).
}\end{minipage}
\label{f_2}
\end{figure}

\section{Hadron Spectra} 
 
For the reaction Pb(158 AGeV) + Pb, 
transverse momentum spectra at mid-rapidity are available also in 6
centrality bins \cite{Cooper}. However, the raw data analysis in \cite{Cooper}
needs refinements and cross checks\footnote{We are grateful to R. Stock,
P. Seyboth, F. Sikler, and P. Jacobs for explaining us the status of the
data analyses.}. 
We, therefore, mention here only that our preliminary 
analysis along the lines in \cite{bk} seems to point to a larger transverse flow
with increasing centrality.

\section{Intermediate-Mass Dileptons} 

As pointed out in \cite{Gale}, the dilepton spectra can be analyzed in the
same spirit as the hadron multiplicities and hadron momentum spectra, 
i.e., one discards any detail of the dynamics and asks only for a simple
parameterization. As a result, one gets 
for the thermal dilepton spectrum
\cite{Gale}
\begin{equation}
\frac{dN}{d^4 Q} = 
\frac{5 \alpha^2}{36 \pi^4} N_{\rm dil}
\exp \left\{ - 
\frac{M_\perp \cosh (Y - Y_{\rm cms} )}{T_{\rm dil}} \right\},
\end{equation}
where $Q$ is the lepton pair's four-momentum,
$M_\perp$ its transverse mass and $Y$ its rapidity;
$Y_{\rm cms}$ denotes the fire ball rapidity, and $ N_{\rm dil}$ is a
normalization factor characterizing the space-time volume of the fire
ball;
flow effects are negligible for invariant mass spectra. 
In \cite{Gale,Gallmeister} we have shown that 
the space-time averaged temperature parameter 
$T_{\rm dil} \approx 170$ MeV
(i.e., a value coinciding with the chemical freeze-out temperature)
provides a \underline{common} description of the low-mass CERES data and the 
intermediate-mass NA50 data at $\sqrt{s} = 17$ AGeV;
also the WA98 photon data are consistent with this value.
It was, therefore, a surprise to us that our analysis of the efficiency 
corrected and centrality binned NA50 data \cite{Capelli}
\underline{separately}
give as optimum fit a temperature scale in the order of 
250 MeV,\footnote{$T_{\rm dil}$ in bins 1 and 2 (very peripheral
collisions) deviates from this value. The extracted values 
are not safe and depend
sensitively on the normalization of the Drell-Yan yield.} 
see figs.~3 and 4. At this temperature the spectral shapes of thermal
dileptons and dileptons from correlated decays of open charm mesons
are nearly identical within the NA50 acceptance.

We have also checked the temperature sensitivity of the parameterization
eq.~(2) by analyzing the new low-mass dilepton
CERES data \cite{CERES} for the reaction
Pb(40 AGeV) + Au and find, when using as important ingredient the
hadronic cocktail, $T_{\rm dil} = 145$ MeV
(which also coincides with the chemical freeze-out temperature
deduced for this beam energy), see fig.~5.

\begin{figure}
\centering
~\\[-.3cm]
\includegraphics[width=12cm,angle=-90]{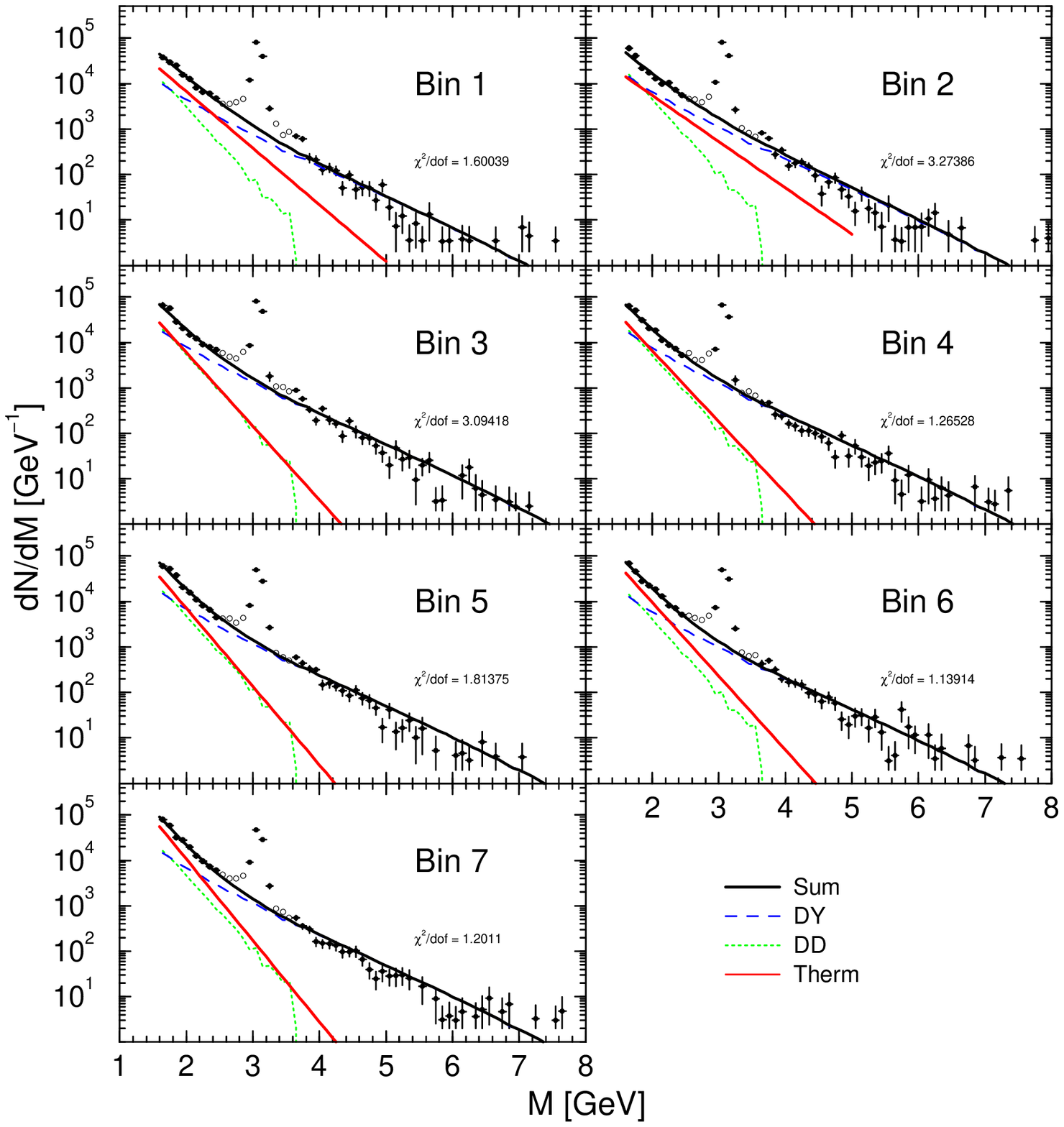}
~\\[.5cm]
\begin{minipage}[t]{14cm}
{\footnotesize Fig.~3:
Fits of the thermal contribution according to eq.~(2) 
to the NA50 data \cite{Capelli} in
7 centrality bins ($E_T = 19, 36, 52, 67, 80, 93, 110$ GeV
for bin 1 - 7).
The Drell-Yan yield and the open charm contributions are calculated
as described in detail in \cite{Gale,Gallmeister}. 
}\end{minipage}
\label{f_3}
\vfill
\centering
~\\[-.1cm]
\includegraphics[width=4cm,angle=-90]{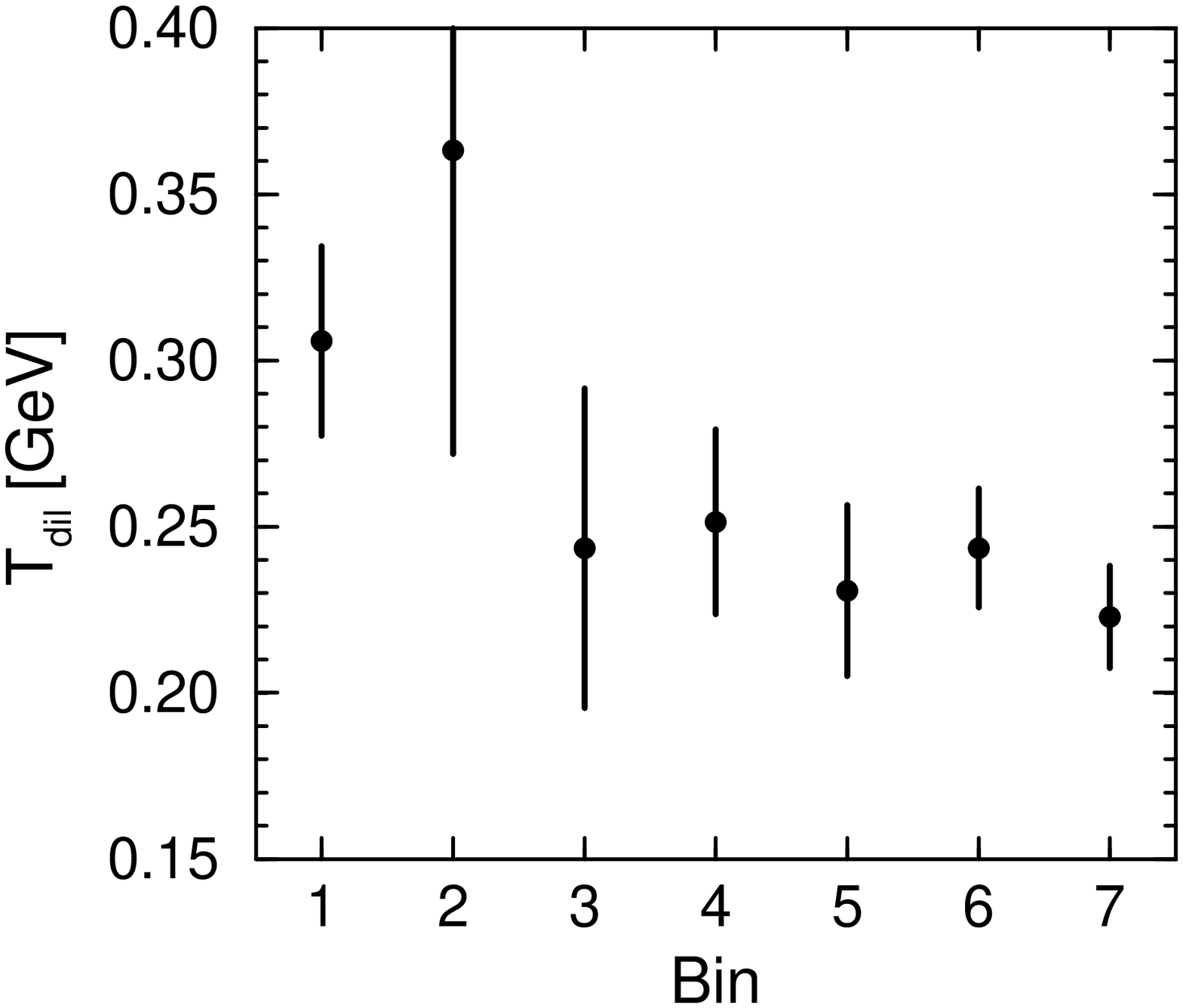}
\hspace*{6mm}
\includegraphics[width=4cm,angle=-90]{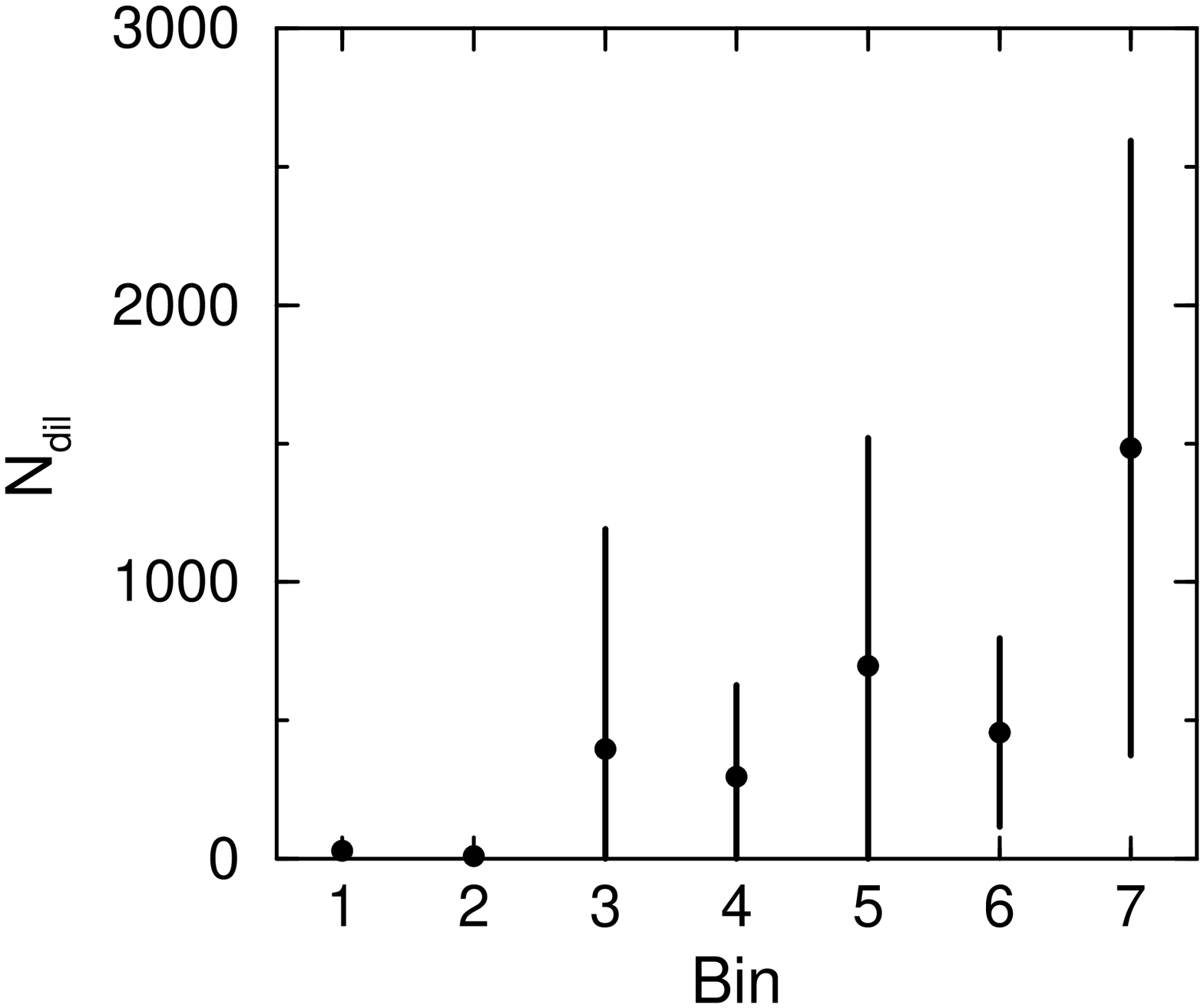}
~\\[.5cm]
\begin{minipage}[t]{14cm}
{\footnotesize \centerline{Fig.~4:
Temperature and space-time normalization factor used in fig.~3.}
}\end{minipage}
\label{f_4}
\end{figure}

\begin{figure}
\begin{minipage}[t]{14cm}
~\center
\begin{minipage}[t]{6cm}
~\vskip -8mm
\includegraphics[width=5cm,angle=-90]{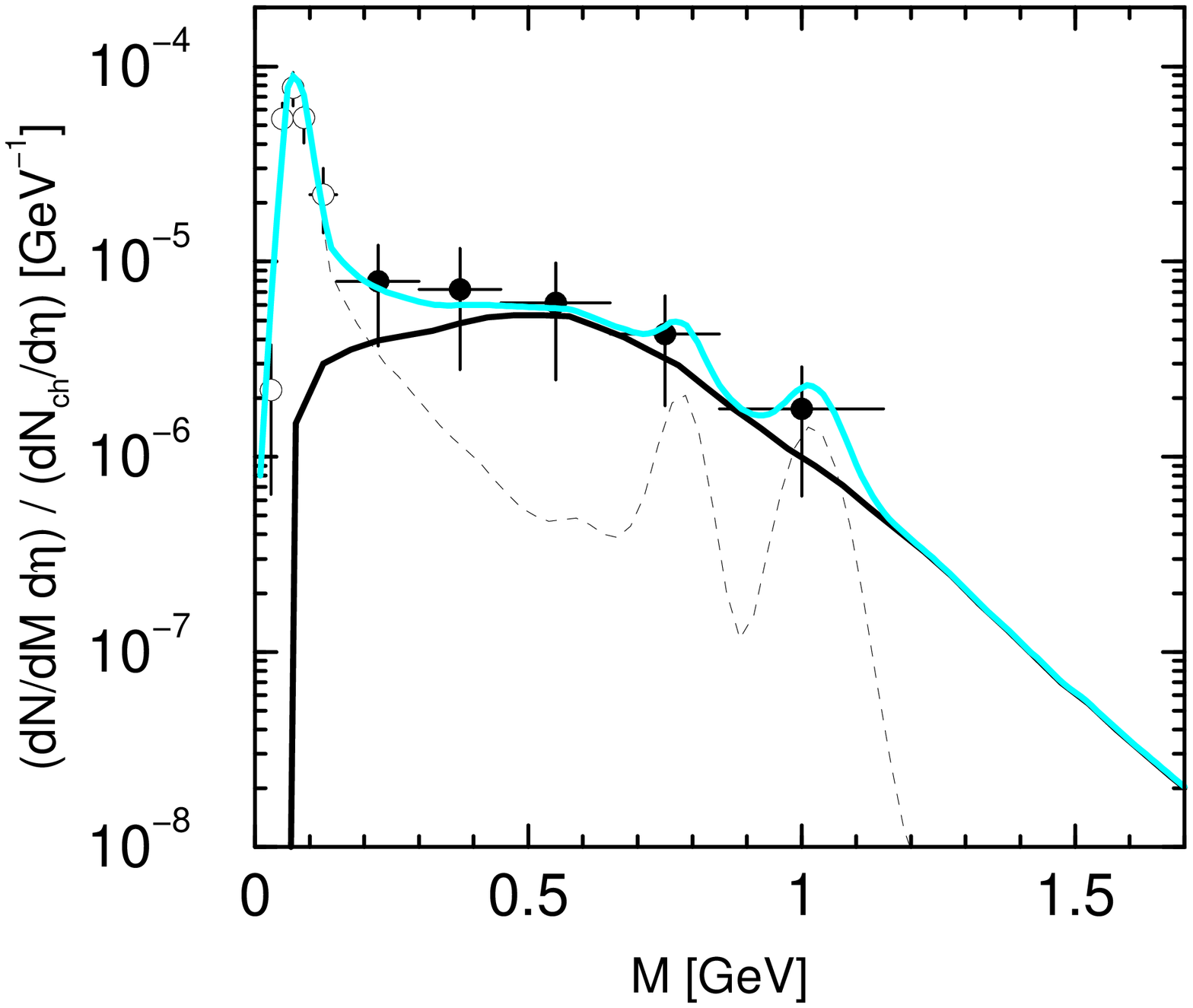}
\end{minipage}
\hspace*{3mm}
\begin{minipage}[t]{6cm}
{\footnotesize Fig.~5:
Comparison of the parameterization eq.~(2) (solid black curve) with the CERES
data \cite{CERES}
for the reaction Pb (40 AGeV) + Au. The thin dashed line 
is the hadronic cocktail,
the gray curve shows the sum.
}\end{minipage}
\end{minipage}
~\vskip -6mm
\label{f_5}
\end{figure}

\section{Summary} 

The analyses of hadron multiplicities, hadron momentum spectra and
inter\-mediate-mass dilepton spectra point to a
centrality independence of the chemical
freeze-out temperature and baryo-chemical potential for 
$\sqrt{s} = 17$ AGeV, 
while the strangeness saturation increases with centrality
for both $\sqrt{s} = 17$ AGeV and 130 AGeV. 
The very preliminary data on transverse
momentum spectra of hadrons indicate a stronger flow in central events
at $\sqrt{s} = 17$ AGeV.
The NA50 data indicate that
the space-time averaged temperature determining the shape of dilepton spectra
stays fairly independent of centrality.


\begin{thebibliography}{99} 
\bibitem{abundance} 
P.~Braun-Munzinger et al., 
Phys.~Lett.~{\bf B344} (1995) 43, 
{\bf B365} (1996) 1, 
{\bf B465} (1999) 15, 
{\bf B518} (2001) 415\\
J.~Cleymans, K.~Redlich, Phys.~Rev.~Lett.~{\bf 81} (1998) 5284
\bibitem{Becattini} F.~Becattini, J.~Cleymans, A.~Keranen, E.~Suhonen,
K.~Redlich,\\
Phys.~Rev. ~{\bf C64} (2001) 024901 
\bibitem{Sikler} F.~Sikler (NA49 collaboration), 
Nucl.~Phys.~{\bf A661} (1999) 45c
\bibitem{Blume} V.~Friese (NA49 collaboration), 
Nucl.~Phys.~{\bf A698} (2002) 487c 
\bibitem{Spencer} J.~Cleymans, B.~K\"ampfer, S.~Wheaton, 
Phys.~Rev.~{\bf C65} (2002) 027901
\bibitem{PHENIX} K.~Adcox et al. (PHENIX collaboration), 
nucl-ex/0112006
\bibitem{Cooper} G. Cooper, Ph.~D. thesis, Berkeley 2000, unpublished
\bibitem{bk} B. K\"ampfer, hep-ph/9612336\\  
B.~K\"ampfer, A.~Peshier, O.P.~Pavlenko, M.~Hentschel, G.~Soff,\\
J.~Phys.~{\bf G23} (1997) 2001 
\bibitem{Capelli} L.~Capelli (NA50 collaboration), Ph.\ D.\ thesis,
University of Lyon, 2000
\bibitem{Gale} K.~Gallmeister, B.~K\"ampfer, O.P.~Pavlenko, C.~Gale,
Nucl.~Phys.~{\bf A688} (2001) 939, {\bf A698} (2002) 424c  
\bibitem{Gallmeister} K.~Gallmeister, B.~K\"ampfer, O.P.~Pavlenko,
Phys.~Lett.~{\bf B473} (2000) 20, 
Phys.~Rev.~{\bf C62} (2000) 057901 
\bibitem{CERES} K. Filimonov et al. (CERES collaboration),
nucl-ex/0109017,\\
S. Damjanovic et al. (CERES collaboration), nucl-ex/0111009 
\end{thebibliography}
\end{document}